# Time synchronization system of Baikal-GVD

V.A. Allakhverdyan[1], A.D. Avrorin[2], A.V. Avrorin[2], V.M. Aynutdinov[2], R. Bannasch[3],
Z. Bardačová[4], I.A. Belolaptikov[1], I.V. Borina[1], V.B. Brudanin[1], N.M. Budnev[5],
V.Y. Dik[1], G.V. Domogatsky[2], A.A. Doroshenko[2], R. Dvornický[1,4], A.N. Dyachok[5],
Zh.-A.M. Dzhilkibaev[2], E. Eckerová[4], T.V. Elzhov[1], L. Fajt[6], S.V. Fialkovsky[7], A.R. Gafarov[5],
K.V. Golubkov[2], N.S. Gorshkov[1], T.I. Gress[5], M.S. Katulin[1], K.G. Kebkal[3], O.G. Kebkal[3],
E.V. Khramov[1], M.M. Kolbin[1], K.V. Konischev[1], K.A. Kopański[8], A.V. Korobchenko[1],
A.P. Koshechkin[2], V.A. Kozhin[9], M.V. Kruglov[1], M.K. Kryukov[2], V.F. Kulepov[7], Pa. Malecki[8],
Y.M. Malyshkin[1], M.B. Milenin[2], R.R. Mirgazov[5], D.V. Naumov[1], V. Nazari[1], W. Noga[8],
D.P. Petukhov[2*], E.N. Pliskovsky[1], M.I. Rozanov[10], V.D. Rushay[1], E.V. Ryabov[5],
G.B. Safronov[2], B.A. Shaybonov[1], M.D. Shelepov[2], F. Šimkovic[1,4,6], A.E. Sirenko[1],
A.V. Skurikhin[9], A.G. Solovjev[1], M.N. Sorokovikov[1], I. Štekl[6], A.P. Stromakov[2],
E.O. Sushenok[1], O.V. Suvorova[2], V.A. Tabolenko[5], B.A. Tarashansky[5], Y.V. Yablokova[1],
S.A. Yakovlev[3], D.N. Zaborov[2]

[1] *Joint Institute for Nuclear Research, Dubna, Russia, 141980*
[2] *Institute for Nuclear Research, Russian Academy of Sciences, Moscow, Russia, 117312*
[3] *EvoLogics GmbH, Berlin, Germany, 13355*
[4] *Comenius University, Bratislava, Slovakia, 81499*
[5] *Irkutsk State University, Irkutsk, Russia, 664003*
[6] *Czech Technical University in Prague, Prague, Czech Republic, 16000*
[7] *Nizhny Novgorod State Technical University, Nizhny Novgorod, Russia, 603950*
[8] *Institute of Nuclear Physics of Polish Academy of Sciences (IFJ PAN), Krakow, Poland, 60179*
[9] *Skobeltsyn Institute of Nuclear Physics MSU, Moscow, Russia, 119991*
[10] *St. Petersburg State Marine Technical University, St. Petersburg, Russia, 190008*

\* E-MAIL: dpetukhov@inr.ru

ABSTRACT: The Baikal-GVD neutrino telescope currently consists of 8 clusters of 288 optical modules (photodetectors). One cluster comprises 8 strings, each of which is subdivided into 3 sections of 12 optical modules. This paper presents the methods of time synchronization between the different GVD components (optical modules, sections, clusters) and estimations of time synchronization accuracy.

# Contents

## 1. Introduction

Currently, the construction of the Baikal-GVD neutrino telescope is underway in Lake Baikal[1][2].To register the Cherenkov radiation of charged particles the optical modules (OM) are used. The photosensitive element of OM is the PMT R7081-100 [3][4]. 12 OMs are combined to the section, which is the main structural element of the Baikal-GVD data acquisition system[5]. Three sections placed on a supporting vertical cable form a string. Eight strings are linked to the single cluster. In total, 8 clusters contain 2304 OM. Each cluster is a functionally complete detector that can register events (muons and cascade showers), either as a stand-alone detector or jointly with other clusters.

We need to estimate the accuracy of time measurements inside GVD because of it directly affects track and cascade reconstruction accuracy. Figure 1 shows how the angular resolution, using the single-cluster reconstruction, depends on muon energy for several different assumptions about the time measurement accuracy. 2 ns is a jitter of photomultiplier and an additional element is a channel intercalibration accuracy of GVD.

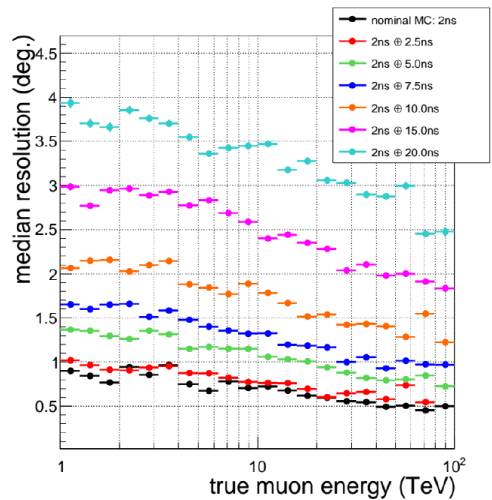

**Figure 1.** Dependence of the angular resolution of the Baikal-GVD cluster (using a χ2-based track reconstruction algorithm) on the accuracy of the time measurement by the telescope channels.

If channel intercalibration accuracy is about 2.5 ns the angular precision of the muon track reconstruction with length more than 300 meters inside the detector is better than 1 degree.

A synchronization of channels inside a single cluster is performed by the common trigger system. For binding the time of GVD event to world time and inter-cluster time synchronization two separate systems are used:
- Synchronization System of Baikal neutrino Telescope (SSBT)
- White Rabbit (WR) from Seven Solutions [6][7]

World time synchronization is made by GMR-5000 Time Server with options of Rubidium oscillator with time drift ±1 ms/year and GPS/GLONASS receiver with binding to world time in interval of ±15 ns.

## 2. Time synchronization system

Figure 2 shows a diagram of the organization of the Baikal-GVD synchronization system.

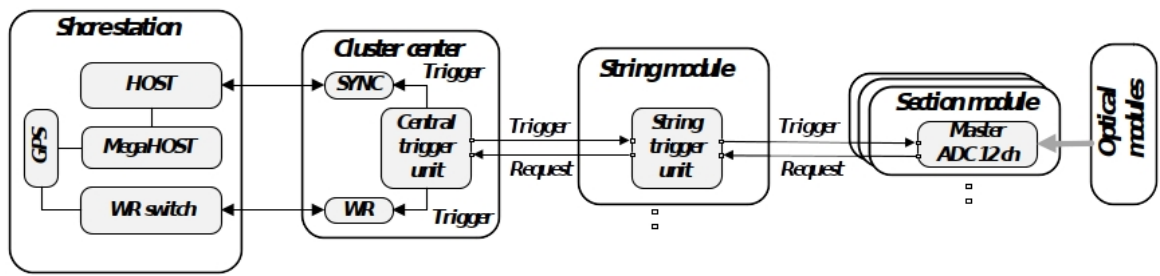

**Figure 2.** Block diagram of the Baikal-GVD synchronization system.

Signals from the 12 optical modules of each section are fed to the ADC unit of the *Master*, forming 12 measuring channels. The *Master* generates a *section request* signal in the case of the coincidence of signals from two adjacent OMs in a time window of 100 ns. The string trigger module generates a *string request* based on *section requests* from belonging sections. Central trigger module of the cluster receives s*tring requests* and forms a *common trigger*. The common trigger is passed to string modules where it branches into three sections. The trigger signal initiates the process of preparing the time frames of the measuring channels, which contain information about the shape of the pulses of the optical modules. A set of time frames from the sections of the cluster initiated by common trigger makes up a physical event. In addition the common trigger goes to both SSBT module and WR module to set world time mark of the event. The accuracy of channels synchronization is determined by the specific of *request* and *trigger* generation by the *Master* unit. The leading edge of these signals is tied to the clock frequency of the electronics units (200 MHz), which leads to a 5-nanosecond uncertainty of their formation time. This does not introduce errors when measuring the time difference of the channels of one section, but affects the registration of pulses by different sections of the cluster. The *trigger units* of the strings (Figure 2) do not cause additional time uncertainty, since for them the time of output signals is tied to the input one. Taking into account the specificities of the functioning of the trigger system, it is possible to estimate the expected accuracy of measuring the time difference Δt by the cluster channels. For channels of one section, the measuring accuracy Δt must be less than 1 ns. This accuracy is achieved by interpolating the pulse shape measured by the ADC with a sampling rate of 5 ns. For channels of different sections, the error Δt consists of the time uncertainties of the moments when the trigger signals are fixed by the *Master* unit of the section. The expected standard deviation of Δt is ~2 ns.

## 3. Study of synchronization accuracy

The accuracy of time measurements can be estimated by detecting the time difference of high energy pulses (>10 single photoelectron levels (SPE)) on separate channels by processing GVD laser calibration runs. Lasers are installed between clusters for calibration of their channels.

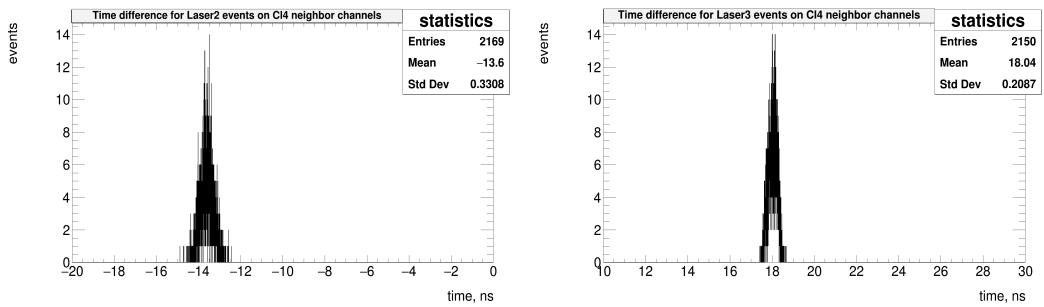

**Figure 3.** Time difference distribution for laser2 (left) and laser3 (right) event detection on neighbor channels of the same sections for cluster 4.

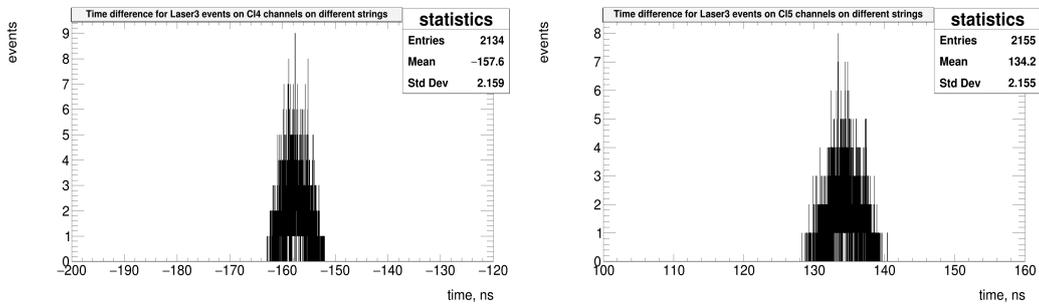

**Figure 4.** Time difference distribution for event detection from laser 3 on channels of different strings for cluster 4 (left) and cluster 5 (right).

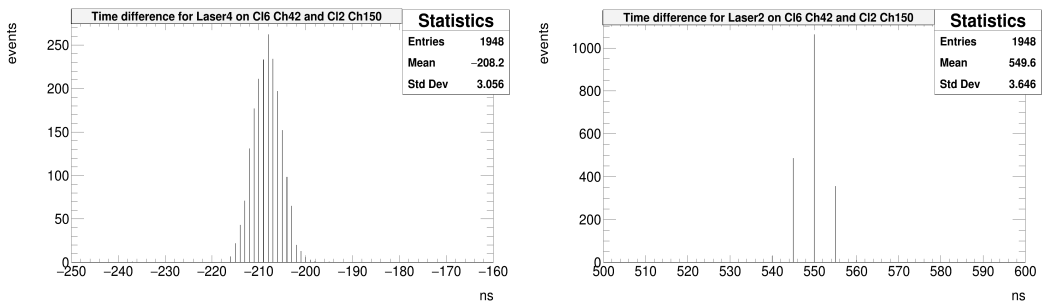

**Figure 5.** Time difference distribution for laser event detection of cluster 2 and cluster 6 for WR (left) and SSBT (right) time synchronization systems.

The value of the standard deviation (RMS) of the measured difference in the response times of channel pairs Δt was used as a parameter characterizing the synchronization accuracy.

The typical results of evaluation the accuracy of the time synchronization inside single cluster are presented on Figure 3 and 4. Estimation of the time synchronization accuracy between clusters is presented on Figure 5.

## 4. Conclusion

Studies of the Baikal-GVD synchronization system with a laser calibration light source showed the correctness of its operation and allowed us to evaluate the synchronization accuracy for all structural elements of the telescope data acquisition system. The results obtained are in good agreement with the expectation. The accuracy of channel synchronization within a single telescope section is significantly better than 1 ns. For different sections of the cluster, this accuracy is about 2 ns. At this level of accuracy, the cluster synchronization system has a minor impact on the angular resolution of the Baikal-GVD cluster. The results of measurements of the synchronization accuracy of two clusters performed by two independent systems are consistent with each other and give the accuracy of events synchronization at the level of 4 ns.

## Acknowledgments

This work was supported by the Ministry of Science and Higher Education of Russian Federation within the financing program of large scientific projects of the "Science" National Project (grant no. 075-15-2020-778).

## References

[1] I.A. Belolaptikov et al. (Baikal-GVD Collaboration), *The Baikal underwater neutrino telescope: Design, performance, and first results*, Astropart. Phys. **7** (1997) 263.

[2] A.V. Avrorin et al. (Baikal-GVD Collaboration), *The Gigaton volume detector in lake Baikal*, Nucl. Instr. and Meth. in Phys. Res. A, **639** (2011) 30.

[3] A.D. Avrorin et al. (Baikal-GVD Collaboration), *The optical module of Baikal-GVD*, EPJ Web of Conferences, **116** (2016) 01003.

[4] A.D. Avrorin et al. (Baikal-GVD Collaboration), *The optical detection unit for Baikal-GVD neutrino telescope*, EPJ Web of Conferences, **121** (2016) 05008.

[5] V.M. Aynutdinov et al. (Baikal-GVD Collaboration), *The data acquisition system for Baikal-GVD*, EPJ Web of Conferences, **116** (2016) 5004.

[6] *J. Serrano et al, The White Rabbit Project, Proceedings of ICALEPS 2009, ICALEPCS TUC004* `http://accelconf.web.cern.ch/accelconf/icalepcs2009/papers/tuc004.pdf`

[7] *Brückner and R. Wischnewski, A time stamping TDC for SPEC and ZEN-platforms based on White Rabbit, ICALEPS-2017,* `PoS(ICRC2015)1041.`